\documentclass[aip,reprint]{revtex4-1}
\usepackage{graphicx}
\usepackage{epstopdf}
\usepackage{booktabs}
\usepackage{grffile}
\usepackage{amsmath}

\begin{document}

% Use the \preprint command to place your local institutional report number 
% on the title page in preprint mode.
% Multiple \preprint commands are allowed.
%\preprint{}

\title{Neural network dose models for knowledge-based planning in pancreatic SBRT} %Title of paper

% repeat the \author .. \affiliation  etc. as needed
% \email, \thanks, \homepage, \altaffiliation all apply to the current author.
% Explanatory text should go in the []'s, 
% actual e-mail address or url should go in the {}'s for \email and \homepage.
% Please use the appropriate macro for the type of information

% \affiliation command applies to all authors since the last \affiliation command. 
% The \affiliation command should follow the other information.
\author{Warren G. Campbell}
\affiliation{Department of Radiation Oncology, University of Colorado School of Medicine}
\author{Moyed Miften}
\affiliation{Department of Radiation Oncology, University of Colorado School of Medicine}
\author{Lindsey Olsen}
\affiliation{Department of Radiation Oncology, University of Colorado Memorial Hospital}
\author{Priscilla Stumpf}
\author{Tracey Schefter}
\author{Karyn A. Goodman}
\affiliation{Department of Radiation Oncology, University of Colorado School of Medicine}
\author{Bernard L. Jones}
%\email[]{Your e-mail address}
%\homepage[]{Your web page}
%\thanks{}
\altaffiliation{Author to whom correspondence should be addressed: bernard.jones@ucdenver.edu}
\affiliation{Department of Radiation Oncology, University of Colorado School of Medicine}

% Collaboration name, if desired (requires use of superscriptaddress option in \documentclass). 
% \noaffiliation is required (may also be used with the \author command).
%\collaboration{}
%\noaffiliation

%\date{\today}
\newcommand{\super}[1]{\ensuremath{^{\textrm{{\tiny #1}}}}}
\newcommand{\subsc}[1]{\ensuremath{_{\textrm{{\tiny #1}}}}}
\newcommand{\enm}[1]{\ensuremath{#1}}
\newcommand{\mat}[1]{\overline{\overline{#1}}}

\begin{abstract}
\textbf{Purpose:} Stereotactic body radiation therapy (SBRT) for pancreatic cancer requires a skillful approach to deliver ablative doses to the tumor while limiting dose to the highly sensitive duodenum, stomach, and small bowel.  Here, we develop knowledge-based artificial neural network dose models (ANN-DMs) to predict dose distributions that would be approved by experienced physicians.

\textbf{Methods}: Arc-based SBRT treatment plans for 43 pancreatic cancer patients were planned, delivering 30-33 Gy in five fractions.  Treatments were overseen by one of two physicians with individual treatment approaches, with variations in prescribed dose, target volume delineation, and primary organs-at-risk.  Using dose distributions calculated by a commercial treatment planning system (TPS), physician-approved treatment plans were used to train ANN-DMs that could predict physician-approved dose distributions based on a set of geometric parameters (vary from voxel to voxel) and plan parameters (constant across all voxels for a given patient).  Patient datasets were randomly allocated, with 2/3\super{rds} used for training, and 1/3\super{rd} used for validation.  Differences between TPS and ANN-DM dose distributions were used to evaluate model performance.  ANN-DM design, including neural network structure and parameter choices, were evaluated to optimize dose model performance.

\textbf{Results}: Remarkable improvements in ANN-DM accuracy (i.e., from $>$30\% to $<$5\% mean absolute dose error, relative to the prescribed dose) were achieved by training separate dose models for the treatment style of each physician.  Increased neural network complexity (i.e., more layers, more neurons per layer) did not improve dose model accuracy.  Mean dose errors were less than 5\% at all distances from the PTV, and mean absolute dose errors were on the order of 5\%, but no more than 10\%.  Dose-volume histogram errors (in cm\super{3}) demonstrated good model performance above 25 Gy, but much larger errors were seen at lower doses.

\textbf{Conclusions}: ANN-DM dose distributions showed excellent overall agreement with TPS dose distributions, and accuracy was substantially improved when each physician’s treatment approach was taken into account by training their own dedicated models.  In this manner, one could feasibly train ANN-DMs that could predict the dose distribution desired by a given physician for a given treatment site.

\textit{This manuscript was submitted to Medical Physics}

\end{abstract}

\pacs{}% insert suggested PACS numbers in braces on next line

\maketitle %\maketitle must follow title, authors, abstract and \pacs

% Body of paper goes here. Use proper sectioning commands. 
% References should be done using the \cite, \ref, and \label commands
%\section{}
%\label{}
%\subsection{}
%\subsubsection{}

% If in two-column mode, this environment will change to single-column format so that long equations can be displayed. 
% Use only when necessary.
%\begin{widetext}
%$$\mbox{put long equation here}$$
%\end{widetext}

% Figures should be put into the text as floats. 
% Use the graphics or graphicx packages (distributed with LaTeX2e).
% See the LaTeX Graphics Companion by Michel Goosens, Sebastian Rahtz, and Frank Mittelbach for examples. 
%
% Here is an example of the general form of a figure:
% Fill in the caption in the braces of the \caption{} command. 
% Put the label that you will use with \ref{} command in the braces of the \label{} command.
%

\section{Introduction}

Pancreatic cancer is a devastating disease with an extremely high mortality.  Over the past decades, treatment for early-to-mid stage pancreatic cancer has evolved significantly.  The best results are seen in patients who are able to undergo surgery, with five-year survival rates of 20-25\% and 4-6\% with and without surgery, respectively [1, 2].  Recently, stereotactic body radiation therapy (SBRT) has emerged as a favorable option for patients with locally advanced or borderline resectable pancreatic adenocarcinoma [3–8].  SBRT is an aggressive local therapy that has improved outcomes in other hard-to-treat tumors, such as non-small cell lung cancer, melanoma, and renal cell carcinoma [9, 10].  By delivering large, ablative doses of radiation in only a few treatment fractions, SBRT leads to significantly improved local control [11].  However, much like surgical techniques, SBRT is a challenging form of local therapy that requires precision to achieve favorable outcomes.

Pancreatic cancer patients who undergo surgical resection of their tumor typically receive a pancreaticoduodenectomy.  This procedure involves the surgical resection of the head of the pancreas, the duodenum, the gallbladder, and often the distal portion of the stomach.  The remaining anatomy is then reassembled to allow bile from the liver and digestive enzymes from the residual pancreas to drain into what remains of the small bowel, helping the patient retain some digestive function.

Our ability to safely perform this surgery today is a result of more than a century of work.  Often referred to as the Whipple procedure, the origins of the modern day pancreaticoduodenectomy are often traced back to a seminal 1935 paper by Whipple, Parsons, and Mullins, in which they presented the procedure as it was performed on three patients [12].  However, the first reported pancreaticoduodenectomy for pancreatic cancer was performed nearly four decades prior in 1898 by Alessandro Codivilla for a case in which the patient died 21 days later from cachexia [13].  In subsequent years, important developments were made before Whipple would refine the procedure, including insights into duodenal function and drainage within the gastrointestinal system [14].  Still, at the time of Whipple’s death in 1963, the surgery remained controversial due to its high morbidity and mortality rates [15, 16].  A 1966 review by Morris and Nardi was more hopeful, emphasizing that if mortality rates for the operation (20-40\% at the time) could be reduced, then “a more optimistic picture” of the procedure could be painted [17].  Today, after decades of refinement, pancreaticoduodenectomy mortality rates have reduced substantially (as low as 1\%) [18].  Even so, expertise and experience still play an important role.  Mortality rates are significantly higher at low-volume facilities compared to high-volume facilities [19].  As such, significant value must still be attributed to the skills of the surgeon.

Here, we argue that radiation therapy’s role in the treatment of pancreatic cancer is following a path parallel to the path for the pancreaticoduodenectomy.  Akin to broad improvements to surgical procedures in general, techniques and technologies involved in radiation delivery have advanced rapidly in recent decades.  Yet, there is still significant variability between treatments at different centers.  Abrams et al recently examined the role that adherence to radiation therapy protocol played in outcomes for the Radiation Therapy Oncology Group (RTOG) 9704 phase III trial for pancreatic cancer, and they found that failure to adhere to protocol was significantly associated with reduced median survival [20].  In theory, inverse planning should result in treatment with an optimal plan, yet studies have shown significant variations between planners [21].  Also, similar to surgery, Amini et al found significant differences in survival for complex radiotherapy to the anal canal between high and low-volume centers [22].

One avenue for refining our approach to pancreatic SBRT is knowledge-based planning, which seeks to utilize information gained from prior radiation therapy treatment plans, including the specific challenges of a given anatomical location, to help guide and improve the radiation therapy planning process.  Currently, most published data regarding the efficacy of pancreatic SBRT come from single-institution studies [4-8].  In an ideal world, it would be possible for centers with experience with this technique to disseminate data regarding how to achieve the best results.  However, in practice, this is often challenging.  Recently, Shiraishi and Moore developed a 3D dose-prediction model that uses artificial neural networks to predict the resulting dose distribution based on patient anatomy [23].  In this paper, we develop a similar dose-prediction model for pancreatic SBRT, for which 3D dose-prediction is important due to the close proximity to highly sensitive organs at risk.

The purpose of this work is to test the feasibility of using knowledge-based planning for 3D dose-prediction in pancreatic SBRT.  We develop artificial neural network dose models (ANN-DMs) to calculate desirable dose distributions for pancreatic cancer patients receiving arc-based SBRT.  Models calculate dose according to geometric and plan parameters, and each model was trained using previous treatment plans and dose distributions at our institution.  To demonstrate the appropriateness of these models, we also developed models of increased or decreased complexity.  In order to guide parameter selection in future models, we also evaluated the relative importance of different parameters within the model.  Ultimately, we are not advocating that the treatment approaches prescribed in this work are the best approaches for pancreatic SBRT.  Rather, we intend to build a framework for the objective comparison of pancreatic SBRT plans that are meant to comply with an established set of treatment guidelines.

\section{Methods}

% Table generated by Excel2LaTeX from sheet 'Sheet2'
\begin{table*}[htbp]
  \centering
    \begin{tabular}{p{6em}|p{32em}|p{15em}}
    \multicolumn{3}{p{49.715em}}{\textbf{Group A}} \\
\hline
    \textbf{Organ} & \textbf{Contouring Guidelines} & \textbf{Constraint} \\
\hline
    Stomach* & Includes: cardia (begins at GEJ), fundus (most cephalad, abuts left hemi-diaphragm, left \& superior to cardia), body (central, largest portion), antrum (gateway to the pylorus). & Stomach minus PTV: Max point dose \enm{<} 30 Gy \\
    Bowel Bag & Loops of small and large bowel and interdigitating mesentery delineated on axial CT slices. Excludes bone, muscle, separate abdominal organs (i.e., kidney, stomach, liver). Includes duodenum. & Bowel minus PTV: Max point dose \enm{<} 30 Gy; V20 \enm{<} 50 cm\super{3} \\
    Spinal Cord & Contoured based on the bony confines of the spinal cord. & Max point dose \enm{<} 10 Gy \\
    Liver* & Gallbladder should be excluded. IVC should be excluded when discrete from liver. PV should be included in liver contour when caudate lobe is seen to left of PV. & Mean dose \enm{<} 10 Gy \\
    Kidneys & Both right and left kidney are contoured in their entirety. & V15 \enm{<} 20\% \\
\hline
    \multicolumn{3}{p{49.715em}}{\textbf{Group B}} \\
\hline
    \textbf{Organ} & \textbf{Contouring Guidelines} & \textbf{Constraint} \\
\hline
    Stomach* & Same as in Group A. & Stomach minus PTV: V33 \enm{<} 1 cm\super{3}, V20 \enm{<} 3 cm\super{3}, V15 \enm{<} 9 cm\super{3} \\
    Duodenum* & \textit{First portion}: begins after pylorus, retroperitoneal after first ~5 cm where it is suspended by hepatoduodenal ligament.  \textit{Second (descending) portion}: starts at superior duodenal flexure, attached to head of pancreas, ~7.5 cm long, located to right of IVC at levels L1-L3.  \textit{Third (transverse) portion}: crosses in from of aorta \& IVC and is posterior to SMA \& SMV, ~10 cm long, marks end of C-loop of duodenum.  \textit{Fourth (ascending) portion}: travels superiorly until it is adjacent to inferior pancreatic body, ~2.5 cm long, lies anteriorly to the IMV until the IMV moves medially at the transition to the jejunum. & Duodenum minus PTV: V33 \enm{<} 1 cm\super{3}, V20 \enm{<} 3 cm\super{3}, V15 \enm{<} 9 cm\super{3} \\
    Small Bowel & Loops of small bowel initiating at the jejunum (end of fourth portion of the duodenum) and extending to the start of the ascending colon. Excludes the duodenum, which is contoured separately. & Small Bowel minus PTV: V33 \enm{<} 1 cm\super{3}, V20 \enm{<} 3 cm\super{3}, V15 \enm{<} 9 cm\super{3} \\
    Large Bowel & Includes ascending, transverse, descending colon loops. Terminates at the rectum. & Large Bowel minus PTV: V33 \enm{<} 5 cm\super{3}, V20 \enm{<} 10 cm\super{3}, V15 \enm{<} 15 cm\super{3} \\
    Spinal Cord & Same as in Group A. & V15 \enm{<} 1 cm\super{3} \\
    Liver* & Same as in Group A. & Liver minus GTV: D50\% \enm{<} 12 Gy \\
    Kidney & Same as in Group A. & V15 \enm{<} 35\% \\
    \end{tabular}%
  \caption{Organs at risk, OAR contouring guidelines, and dose constraints for patients in Group A and Group B.  GEJ: gastroesophageal junction, IVC: inferior vena cava, PV: portal vein, SMA: superior mesenteric artery, SMV: superior mesenteric vein, IMV: inferior mesenteric vein. *Contouring per RTOG guidelines.}
\end{table*}%

\subsection{Patient Data}
Data were collected from 43 patients with locally advanced, borderline resectable, or recurrent pancreatic tumors treated at our institution using arc-based SBRT.  These retrospective data were collected under an internal review board-approved protocol to analyze novel methods in patient dosimetry.  Planning target volumes (PTVs) were formed via a 0-5 mm patient-specific anisotropic expansion from physician-defined clinical target volumes (CTVs).  Median (\enm{\pm\sigma}) PTV was 110 (\enm{\pm}77) cm\super{3}.  Tumors were prescribed to receive maximum doses of 30-33 Gy in five fractions, delivered using 2-4 coplanar arcs spanning 250-360 degrees.  Patient immobilization and repositioning was achieved using Alpha Cradle expanding foam forming molds (Smithers Medical Products, Inc.; North Canton, OH).  Tumor motion was managed using either abdominal compression or respiratory gating.

Patients were treated by one of two physicians, each with an individualized treatment approach.  As such, two distinct groups of patients were identified.  In Group A (29 patients), two treatment volumes were specified: a larger PTV prescribed to receive 20 Gy, and a smaller, gross tumor volume (GTV) prescribed to receive 30 Gy.  In Group B (14 patients), a single treatment volume, the PTV, was prescribed to receive 33 Gy.  Primary organs at risk (OARs) for both Groups A and B are listed in Table I, along with contouring guidelines and dose constraints.  Select OARs were contoured according to specific RTOG guidelines [24].

\subsection{Artificial Neural Network Dose Models}
Artificial neural networks can feature multiple hidden layers between an input layer and an output layer, and each hidden layer can include multiple nodes.  A simple example of an artificial neural network is depicted in Figure 1, which features two inputs, one hidden layer with three nodes, and a single output.  The activation value for each hidden node is determined by taking a weighted sum of the input values (or, for multi-layer networks, the set of nodes in the previous hidden layer) and then entering that weighted sum into an activation function (e.g., a sigmoid function).  After activation values are calculated for each node in a layer, activation values for the subsequent layer (in this case, the output node) are calculated in a similar fashion.  All weight values that interconnects nodes of one layer to nodes of the next layer are independent of one another, and these values are adjusted as the neural network is trained.  Initially, all weight values are chosen randomly.  Then, by providing training data (i.e., input values with known output values), error magnitudes observed at the output layer determine how weights should be adjusted, typically through backpropagation.  This routine is repeated multiple times using known datasets until weight adjustments no longer have discernible effects on error size.  Then, a validation dataset (i.e., input values with known output values that were not used during training) is used to assess the network’s performance.

\begin{figure*}
 \includegraphics{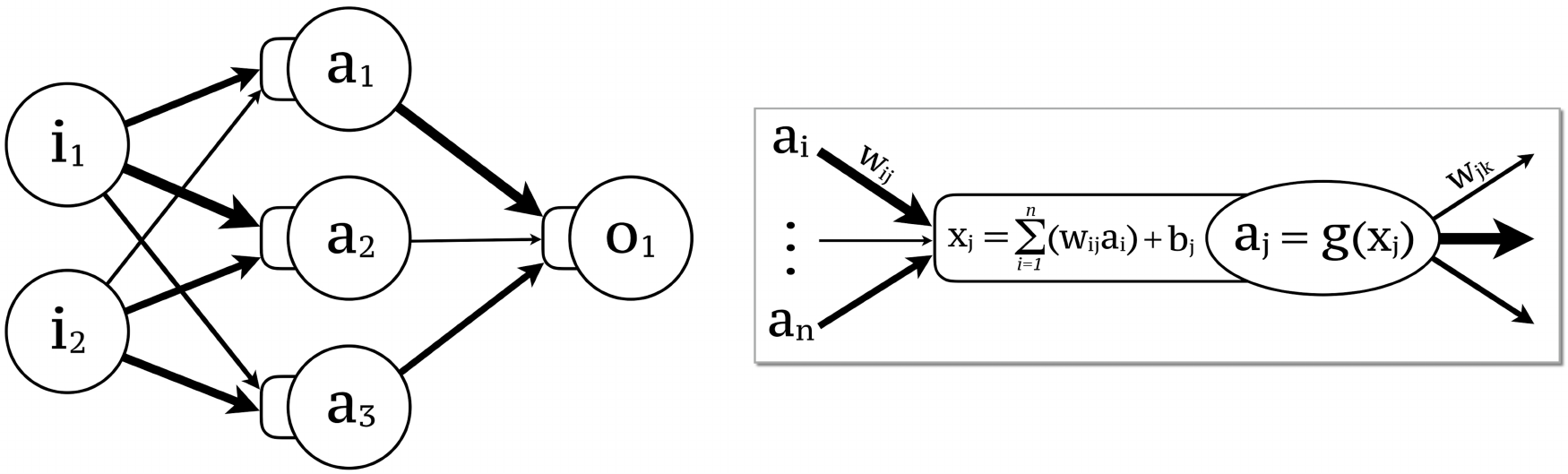}%
 \caption{Left: diagram showing a simple artificial neural network with two inputs, a hidden layer with three nodes, and one output.  Right: a more detailed depiction of how the activation value of a single node, a\subsc{j}, is calculated.  A weighted sum of all activation values from the prior layer plus a bias value is indicated here by x\subsc{j}.  This value is then entered into g, an activation function (e.g. we used log-sigmoid functions).  Inter-node weight values (i.e., w\subsc{ij}, w\subsc{jk}, etc.) are independent of one another, and these values are adjusted as the neural network is trained.}
 \end{figure*}

For each patient, a volume of interest was defined as all voxels within 100 mm of the surface of the PTV.  Each individual voxel in the volume of interest could then be used to provide input data for an artificial neural network that would produce a single output value: dose to that voxel.  Input values were either geometric parameters or plan parameters.  Geometric parameters are factors such as the voxel’s distance to the PTV surface, distance to an OAR, or the number of arcs directly impinging on the voxel.  Plan parameters are factors such as the photon beam energy or PTV volume.  Geometric parameters can differ for each voxel within a given patient, whereas plan parameters are equal for all voxels in a single patient, but can differ between patients.  The initial parameters used for this work are listed in Table II and are described in more detail below.  Ultimately, four main ANN-DMs were developed: a pair of models each for Group A and Group B, with a model for within the treated volume, and a model for outside the treated volume.

%\begin{figure}
 %\includegraphics{fig1_4DCT_trajectories.pdf}%
 %\caption{Projected 3D trajectories of 14 pancreatic tumors in the coronal (left) and sagittal (right) planes, as measured from 4DCT images.  The trajectories show the movement of the tumor centroid, are drawn in approximate relation to one another using the common bile duct as positional reference. }
 %\end{figure}

% Table generated by Excel2LaTeX from sheet 'Sheet1'
\begin{table*}[htbp]
  \centering
    \begin{tabular}{p{9em}p{6em}p{16em}llp{6.5em}}
    \toprule
    \toprule
    \textbf{Parameter} & \textbf{Name} & \textbf{Description} & \multicolumn{1}{p{4.215em}}{\textbf{Min}} & \multicolumn{1}{p{4.215em}}{\textbf{Max}} & \textbf{Units} \\
\hline
    \textbf{All} & \multicolumn{1}{c}{} & \multicolumn{1}{c}{} &       & \multicolumn{2}{r}{} \\
\hline
    \midrule
    ag\subsc{1}   & r\subsc{ptv3D} & Shortest 3D distance to PTV & -24.7 & 100.9 & mm \\
    ag\subsc{2}   & r\subsc{ptv2D} & Axial component of rptv3D & -25   & 100   & mm \\
    ag\subsc{3}   & z\subsc{in}   & Relative Sup-Inf position, in slice & 0     & 1     & normalized \\
    ag\subsc{4}   & z\subsc{out}  & Sup-Inf position, out of slice & 0     & 1     & normalized \\
    ag\subsc{5}   & r\subsc{surf} & Depth from patient surface & 0     & 109.2 & mm \\
    ag\subsc{6}   & F\subsc{arc}  & Arc factor & 0     & 1     & normalized \\
    ag\subsc{7}   & \enm{\theta}     & Angle relative to PTV & -3.14 & 3.13  & radians \\
    ap\subsc{1}   & V\subsc{ptv}  & Target volume & 19.67 & 175.5 & cm\super{3} \\
    ap\subsc{2}   & E     & Photon Energy & 6     & 10    & MV \\
\hline
    \textbf{Group A} & \multicolumn{1}{c}{} & \multicolumn{1}{c}{} &       & \multicolumn{2}{r}{} \\
\hline
    \midrule
    ag\subsc{8}   & r\subsc{st}   & Shortest distance to stomach & -23.6 & 165.5 & mm \\
    ag\subsc{9}   & r\subsc{bb}   & Shortest distance to bowel bag & -34.1 & 105.5 & mm \\
    ag\subsc{10}  & r\subsc{gtv}  & Shortest distance to GTV & -13.7 & 110.5 & mm \\
\hline
    \textbf{Group B} & \multicolumn{1}{c}{} & \multicolumn{1}{c}{} &       & \multicolumn{2}{r}{} \\
\hline
    \midrule
    ag\subsc{8}   & r\subsc{st}   & Shortest distance to stomach & -15.9 & 130.2 & mm \\
    ag\subsc{9}   & r\subsc{duo}  & Shortest distance to duodenum & -12.1 & 121.5 & mm \\
    ag\subsc{10}  & r\subsc{sb}   & Shortest distance to small bowel & -26.1 & 130   & mm \\
    \bottomrule
    \bottomrule
    \end{tabular}%
\caption{Initial parameters used to build the artificial neural network dosimetric models (ANN-DMs).  Geometric parameters ag\subsc{1} through ag\subsc{7} and both plan parameters were common to all ANN-DMs and would be generically relevant for most treatment sites receiving arc-based SBRT.  Three site-specific geometric parameters were also included for patients in Groups A and B.  For all parameters, minimum and maximum values are provided, along with their units.}
\end{table*}

\subsection{Geometric Parameters}
In general, all voxels were categorized into two main regions: (1) “in slice” voxels, which include all voxels that lie in an axial slice that contains at least some portion of the PTV, and (2) “out of slice” voxels, which include all other voxels.

The primary parameter in the geometric model is r\subsc{ptv3D}, the shortest 3D distance from the voxel to the PTV surface.  This factor was meant to capture the general shape of the dose gradient outside the target.  A related parameter, r\subsc{ptv2D}, describes the axial distance from the voxel to the PTV surface (i.e., the shortest distance ignoring the superior-inferior displacement).  Inclusion of this factor allows for more accurate differentiation in the model between voxels inside and outside treated slices (i.e., slices that include at least a portion of the treatment volume).  Parameter z\subsc{in} is the normalized superior-inferior (SI) distance from the voxel to the PTV centroid for voxels within the treated slice (i.e., ranging from 0 for voxels in the central PTV slice to 1 for voxels in the most superior or most inferior PTV slice).  This factor captures photon scatter effects between slices.  Parameter z\subsc{out} is the normalized SI distance to the PTV (normalized relative to PTV height) for voxels out-of-slice with the target, and this helps to further characterize the scatter distribution outside of the direct beam.  The depth parameter, r\subsc{surf}, is the distance from the voxel to the patient surface along the axis between the voxel and the PTV.  The arc factor, F\subsc{arc}, is a binary parameter that describes whether a voxel lies in the direct path of an incoming treatment beam.  Most plans examined in this study used 360$^{\circ}$ arcs, but some plans used anterior arcs that do not directly irradiate posterior regions such as the kidneys or spinal cord.  The parameter \enm{\theta} is the angle within the axial plane, calculated with respect to the PTV centroid.
Three geometric parameters specific to the pancreas were also included in each patient group for better dose model performance.  The majority of these parameters were selected based on the OARs nearest the target volumes.  Both Group A and Group B included r\subsc{st}, the shortest distance between the voxel and the stomach volume.  Group A also included a second OAR factor, r\subsc{bb}, which is the shortest distance between the voxel and the “bowel bag” volume.  An additional factor in Group A, r\subsc{gtv}, describes the shortest distance between the voxel and the GTV, and captures the effect of the multiple dose prescription levels in Group A plans.  Group B patients included two additional OAR factors: r\subsc{duo} and r\subsc{sb}, which respectively are the shortest distances between the voxel and the duodenum volume and small bowel volume.  For all distance-to-volume based geometric parameters, distance values are negative when voxels occur within the volume in question.

\subsection{Plan Parameters}
Two plan parameters were chosen to capture broad differences between plans for different patients.  The first parameter, V\subsc{ptv} , denotes the volume of the PTV in cm\super{3}.  The second parameter, E, indicates the photon energy used for treatment, either 6 MV, 10 MV, or 10 MV in flattening filter free (FFF) mode.  Two additional plan parameters were eliminated when patients were divided into two separate groups.  First eliminated was Rx, which indicated the prescribed maximum dose.  Second eliminated was MD, which indicated the identity of the approving physician.  The factor MD was initially considered in order to account for differences in physician treatment style, including the types of hot spots considered permissible, OAR dose-volume histogram priorities, and relative tradeoffs between PTV coverage and OAR sparing.  Separating patients into Groups A and B effectively incorporates these two factors into the dose models without having to explicitly indicate them for each patient.

\subsection{Model Training and Validation}
Multiple ANN-DMs were developed, but all were trained and validated using data exported from the TPS (Eclipse; Varian Medical Systems; Palo Alto, CA), including CT images, treatment plans, structures, and TPS-calculated dose distributions.  For each individual voxel, inputs to the ANN-DM were geometric and plan parameters (as described above), and the single output was a prediction of the TPS-calculated dose for that voxel.  By inputting thousands of voxels across multiple patients, prediction models are established for each ANN-DM to estimate physician-approved dose.

Patient datasets were randomly divided into two groups: roughly 2/3\super{rds} for training, and 1/3\super{rd} for validation (19 and 10 for Group A, 9 and 5 for Group B, respectively).  Only voxels within 100 mm of the PTV were used as inputs.  Across all 43 patients, a total of 75,610,117 voxels were available for training and validation, with 812,908 of those voxels residing within the PTV.  However, due to the highly correlated nature of TPS dose distributions, the number of truly unique data points is likely smaller.  To lower computation times, a random subset of 1 in every 40 voxels (2.5\% of the total) was used for training.  This choice was validated by also training models with 5\%, 10\%, and 25\% of the total voxels, and no significant changes were seen in the results.  For each group of patients, two ANN-DMs were computed: one for voxels inside the PTV, and another for voxels outside the PTV.  All computations were performed in the MATLAB computing environment (MathWorks; Natick, MA).  To produce a model that was most accurate in the area close to the PTV, voxels were weighted by dose (i.e., weight = dose/prescription) to ensure sufficient sampling of the dose falloff region.

The majority of ANN-DMs used in this study were composed of a feed-forward network with 25 nodes in a single hidden layer, and they were trained using L2 regularization.  Log-sigmoid functions were used for all activation functions, and scaled conjugate gradient backprojection was used to train each network.  Other ANN training algorithms were investigated (i.e., quasi-Newton backprojection and conjugate gradient backprojection), but no significant improvements to the results were seen.  To evaluate the role of neural network complexity, multiple networks with increased and decreased complexity were also tested using a consistent set of model parameters.  Neural network complexity increases with increasing number of hidden layers and increasing number of nodes per layer.  Here, we tested neural networks with 1 to 3 hidden layers, with each layer containing 10 to 50 nodes in each layer.  Mean absolute dose error was used to quantify any benefits gained with increasing complexity.

\subsection{Gaussian Broadening of Target Volumes}

In early efforts to train the models, we observed good performance within the training dataset, but very poor performance in the validation dataset.  These errors were found to originate mainly from overfitting of the volume parameter, V\subsc{ptv}, which is a continuous variable, but takes a discrete value for all voxels belonging to a single patient.  To correct for this, Gaussian noise was added to the V\subsc{ptv} value seen in each voxel for a given patient.  The standard deviation of this distribution was chosen empirically to be 10 cm\super{3}, as this was found to generate nearly continuous coverage of the volume parameter space across all patients.  In this way, we were able to prevent strange model behavior in the validation dataset.

%\begin{figure*}
 %\includegraphics{fig4_SIMotion.pdf}%
 %\caption{Tumor trajectories in the superior-inferior (SI) direction. Each vertical box denotes the distance encompassing 95\% of the daily motion during one fraction of treatment.  Fractions are grouped by patient, and the horizontal bars denote the amplitude of SI motion measured in the 4DCT images for that patient.  In addition to highlighting the tendency of 4DCT to underestimate pancreatic motion, these data also demonstrate the relative stability of motion in a single patient (from day-to-day), as well as the large differences in motion between different patients.  Tumor position and 4DCT amplitude are shown relative to the mean position.}
 %\end{figure*}

\section{Results}

% Table generated by Excel2LaTeX from sheet 'Sheet3'
\begin{table*}[htbp]
  \centering
    \begin{tabular}{p{6.5em}|p{4.215em}l|p{4.215em}l|p{4.215em}l}
    \textbf{Parameter} & \multicolumn{2}{p{8.43em}|}{\textbf{Inside PTV}} & \multicolumn{2}{p{10.785em}|}{\textbf{$\leq$ 3 cm Outside PTV}} & \multicolumn{2}{p{10.36em}}{\textbf{$>$ 3 cm Outside PTV}} \\
\hline
    \multicolumn{1}{l|}{} & \textbf{A} & \multicolumn{1}{p{4.215em}|}{\textbf{B}} & \textbf{A} & \multicolumn{1}{p{6.57em}|}{\textbf{B}} & \textbf{A} & \multicolumn{1}{p{6.145em}}{\textbf{B}} \\
\hline
    \textbf{r\subsc{ptv3D}} & \multicolumn{1}{l}{1.8} & 11.8  & \multicolumn{1}{l}{4.9} & 12.8  & \multicolumn{1}{l}{0.9} & 15.6 \\
    \textbf{r\subsc{ptv2D}} & \multicolumn{1}{l}{3.8} & 5.3   & \multicolumn{1}{l}{13.5} & 14    & \multicolumn{1}{l}{5} & 0.9 \\
    \textbf{z\subsc{in}} & \multicolumn{1}{l}{2.8} & 4.8   & \multicolumn{1}{l}{12.5} & 8.4   & \multicolumn{1}{l}{5.8} & 4.9 \\
    \textbf{z\subsc{out}} & \multicolumn{1}{l}{1.6} & 24.2  & \multicolumn{1}{l}{40.2} & 37.6  & \multicolumn{1}{l}{10.5} & 2.8 \\
    \textbf{r\subsc{surf}} & \multicolumn{1}{l}{7} & 7.7   & \multicolumn{1}{l}{1.7} & 6.9   & \multicolumn{1}{l}{-0.8} & 12.4 \\
    \textbf{F\subsc{arc}} & \multicolumn{1}{l}{0.1} & 5.7   & \multicolumn{1}{l}{8.6} & 1.6   & \multicolumn{1}{l}{3.8} & -7.2 \\
    \textbf{V\subsc{ptv}} & \multicolumn{1}{l}{0} & 11.3  & \multicolumn{1}{l}{6.7} & 3.3   & \multicolumn{1}{l}{1.4} & -5.2 \\
    \textbf{r\subsc{st}} & \multicolumn{1}{l}{2.8} & 4.6   & \multicolumn{1}{l}{1.1} & 8.6   & \multicolumn{1}{l}{-1.4} & 9.7 \\
    \textbf{r\subsc{bb}} & \multicolumn{1}{l}{0.7} & \multicolumn{1}{p{4.215em}|}{--} & \multicolumn{1}{l}{9.9} & \multicolumn{1}{p{6.57em}|}{--} & \multicolumn{1}{l}{10.3} & \multicolumn{1}{p{6.145em}}{--} \\
    \textbf{r\subsc{gtv}} & \multicolumn{1}{l}{31.2} & \multicolumn{1}{p{4.215em}|}{--} & \multicolumn{1}{l}{2.5} & \multicolumn{1}{p{6.57em}|}{--} & \multicolumn{1}{l}{1.2} & \multicolumn{1}{p{6.145em}}{--} \\
    \textbf{r\subsc{duo}} & --    & 8.7   & --    & 0.2   & --    & -4.2 \\
    \textbf{r\subsc{sb}} & --    & -1.2  & --    & 10.8  & --    & 1.3 \\
    \end{tabular}%
\caption{Quantifying the relative contribution of each parameter to model accuracy.  Values shown are the relative increase in mean absolute dose error (as \% of Rx dose) when each parameter was removed from the dose model.  Analysis was divided into three regions: inside the PTV, within 3 cm outside of the PTV, and more than 3 cm outside of the PTV.}
\end{table*}%

\begin{figure}
 \includegraphics{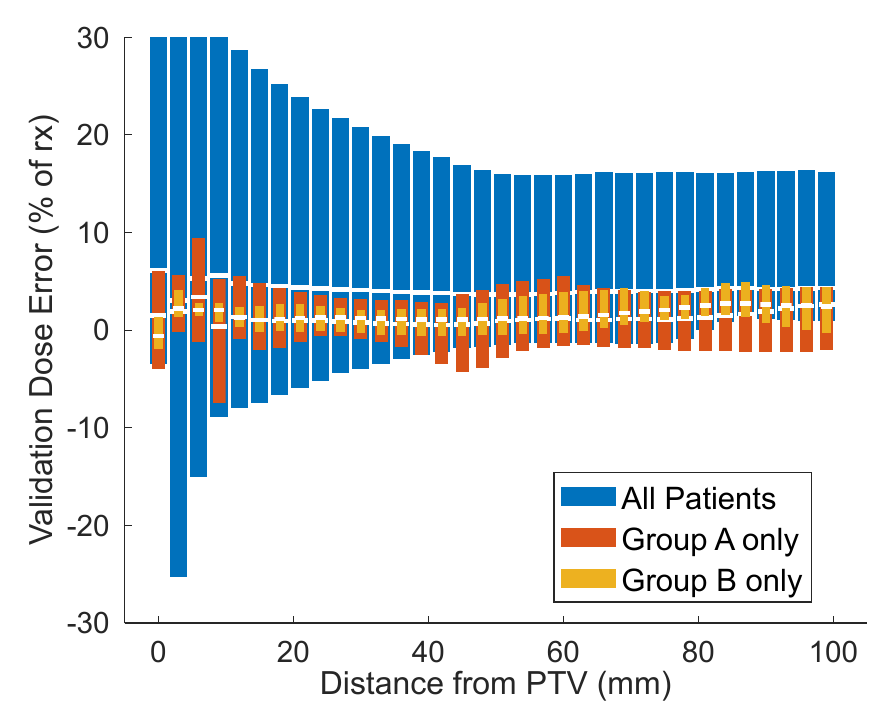}%
 \caption{Substantial reductions in ANN-DM dose error were achieved by creating dedicated models for each distinct group of patients.  Box plots (mean, 95\% CI) of dose errors in validation datasets are shown with respect to distance from the PTV for All Patients, Group A, and Group B.  Most notable in the dose falloff region, mean absolute dose errors were decreased from $>$30\% to $<$5\% when groups were modelled separately.}
 \end{figure}

It became evident early on in ANN-DM development that separate models needed to be created for patients from Group A and Group B in order to account for the distinct treatment approaches taken by each group’s respective physician.  To demonstrate this, Figure 2 illustrates the substantial improvements in accuracy that were gained by creating dedicated models for each group.  Mean absolute dose errors of $>$30\% in the dose falloff region were reduced to $<$5\% when each group was modelled separately.  After preliminary investigations, several different dose model configurations were implemented in order to evaluate choices of ANN-DM structure.  Two main aspects were considered: neural network complexity, and choices of input parameters.

\subsection{Model Complexity}
Mean absolute dose error with respect to model complexity was evaluated by varying the number of hidden layers in the network (1-3) and the number of nodes per layer (10, 20, 30, 40, and 50).  Increased neural network complexity (i.e., more hidden layers and nodes per layer) did not significantly improve model performance.  Although some reduction in errors could be seen in the training dataset as the number of hidden layers increased, those reductions were not reflected in the validation dataset.  Ultimately, a consistent model structure was chosen to include one hidden layers with 25 nodes each.  Using this structure, each ANN-DM took roughly 5 minutes to train on a standard Intel Core-i7 desktop, or roughly 20 seconds on a commercial-grade GPU (Nvidia Quadro M6000).

\begin{figure*}
 \includegraphics{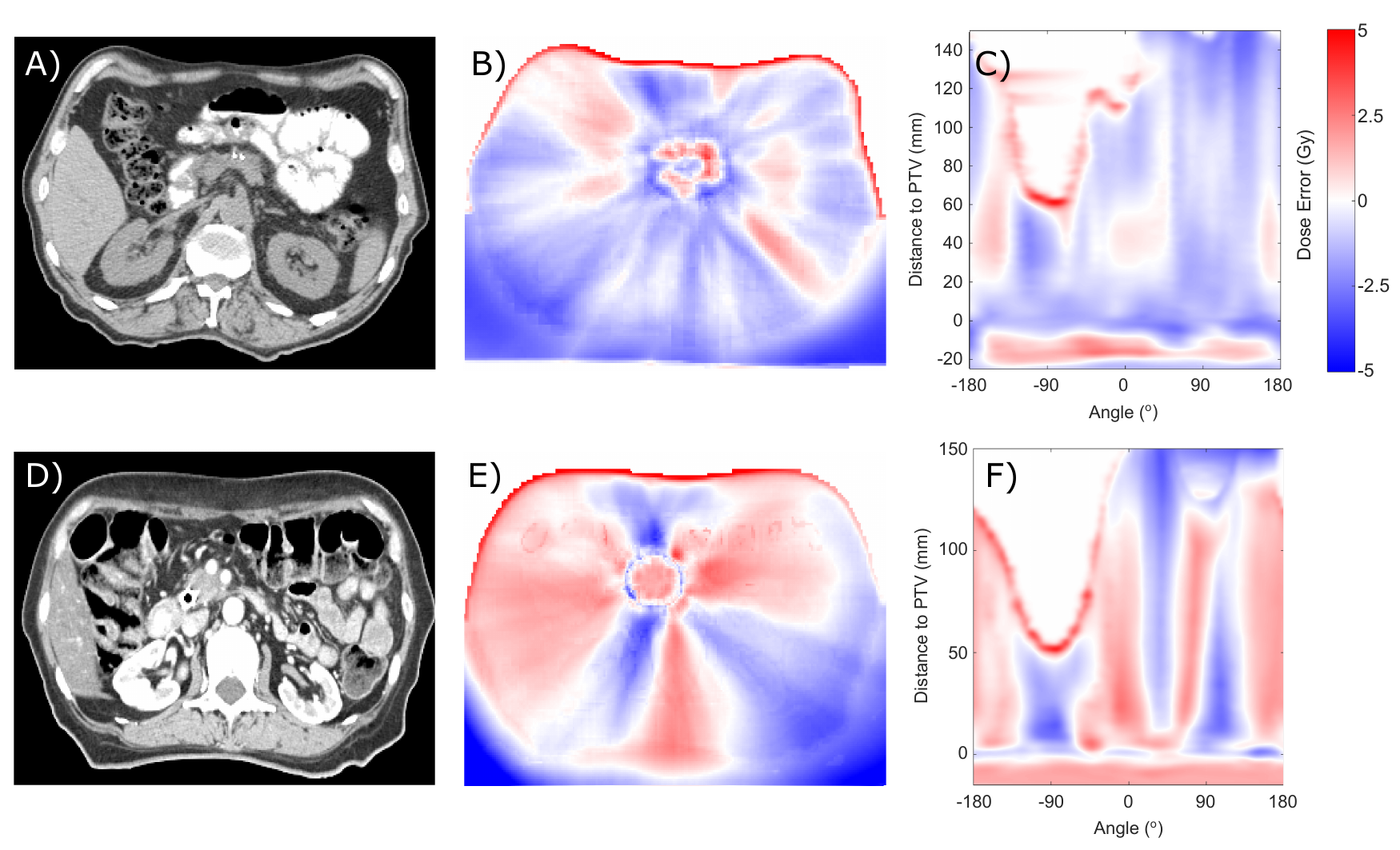}%
 \caption{Sample patient data using final versions of each dose model.  From Group A: (a) an axial slice from the planning CT, (b) the corresponding dose error map, and (c) the dose error map displayed with respect to angle and distance to the PTV centroid.  The same data from Group B is respectively shown in (d), (e), and (f).}
 \end{figure*}

\subsection{Relative Pertinence of Model Parameters}
To quantify the relative impact each parameter had for each dose model, parameters were individually removed from each model to see how much its absence increased that model’s mean absolute dose error.  Dose errors were summarized for three different regions: inside of the PTV, within 3 cm just outside of the PTV, and more than 3 cm outside of the PTV.  Results from these tests are provided in Table III.  The most impactful parameter was z\subsc{out}, which saw a 40\% increase to mean absolute dose error in the region just outside of the PTV.  Nevertheless, all other parameters included in Table III showed considerable pertinence in at least one of the three regions for either Group A or Group B.

Two input parameters from Table II were not included in Table III because it was decided to remove them from all dose models: E and \enm{\theta}.  Removal was either due to the parameter being redundant, or its removal from dose models resulted in improved accuracy in all three regions.  For Group A, removal of the E factor reduced the model’s overall error by roughly 4.5\% of the prescribed dose.  For Group B, all treatment plans used 10 MV FFF beams, so inclusion of the E parameter was unnecessary.  For both groups, inclusion of \enm{\theta} also resulted in increased errors in all regions.
  
\begin{figure*}
 \includegraphics{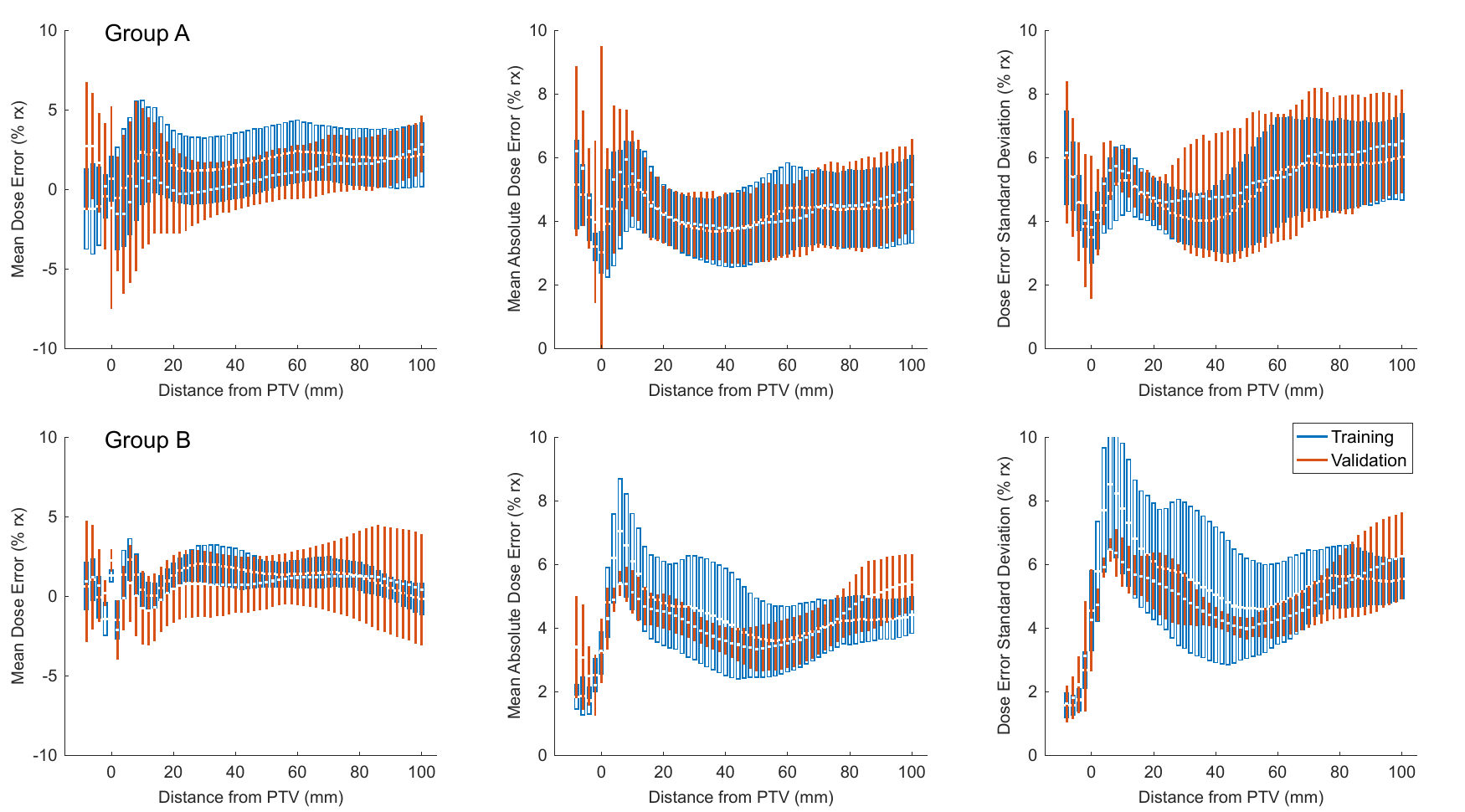}%
 \caption{Dose errors from final versions of each dose model.  Mean dose errors, mean absolute dose errors, and dose error standard deviations are shown separately for Groups A and B, plotted as a function of distance from the PTV.  Voxels just within the boundaries of the PTV (indicated by negative distances from the PTV) are also included to sample the full range of the dose fall off region.  Box plots indicate mean values and 95\% CI ranges.}
 \end{figure*}

\begin{figure*}
 \includegraphics{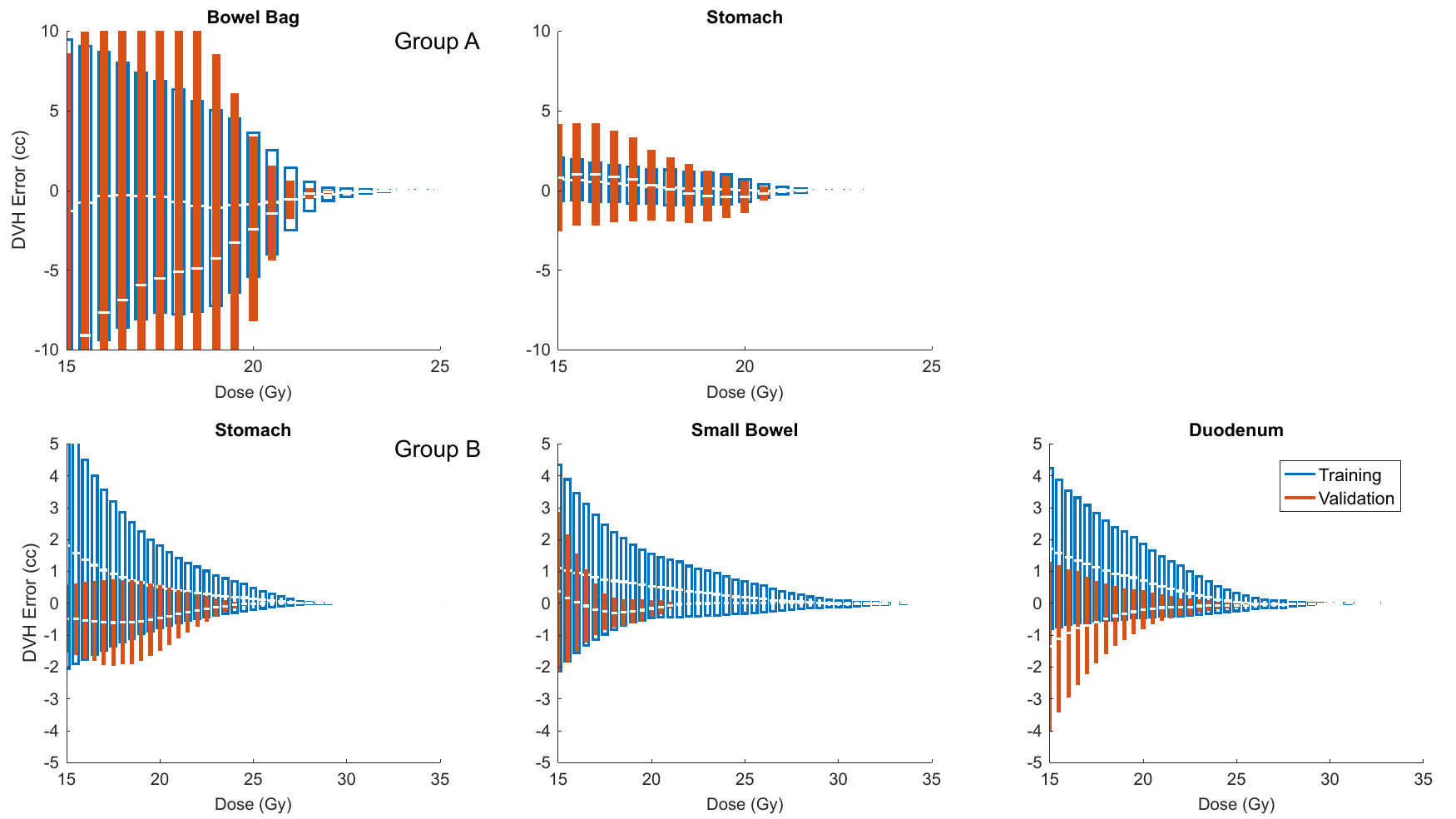}%
 \caption{Each patient’s model-calculated dose distribution was analysed in its entirety to calculate dose-volume histograms (DVHs) for primary organs-at-risk.  Here, DVH errors are shown (in cm\super{3}) for each primary organ-at-risk in Groups A and B.}
 \end{figure*}

\subsection{Model Accuracy}
Using the final versions of each dose model (i.e., with E and \enm{\theta} removed), sample data are shown in Figure 3 for two patients, one from Group A and one from Group B.  Included are axial slices from the planning CT, their corresponding axial dose error maps, as well as dose error maps displayed with respect to angle and distance to the PTV centroid.  Across all patients, the maximum and minimum dose outputs for our ANN-DMs were 36.38 Gy and -1.25 Gy, respectively.

Three performance metrics - mean dose error, mean absolute dose error, and dose error standard deviation - were calculated for training and validation data for Group A and Group B, and are plotted in Figure 4 with respect to voxel distance to the PTV surface.  In order to present the entire dose fall off region, values just within the PTV surface (i.e., negative distance from PTV values) are also shown.  Overall, modelled dose distributions displayed excellent agreement with planned dose distributions.  Mean dose errors were less than 5\% at all distances from the PTV, while mean absolute dose error was on the order of 5\%, but no more than 10\%.  The magnitude of errors were similar between Group A and Group B, although the mean error and mean absolute error were both higher inside the PTV in Group A, possibly due to the multiple dose levels used.  The standard deviation of the dose distribution quantified the distribution of errors, and in both groups was similar in the training and validation datasets.

DVH errors for training and validation data are shown in Figure 5 for each of the primary OAR volumes used for Group A and Group B.  Models displayed good performance in higher dose regions of the DVH curve, but much larger errors were seen at lower doses.  Based on the OAR dose restraints used in this work (see Table I), only the V33 metric could be reliably predicted.
 
\begin{figure*}
 \includegraphics{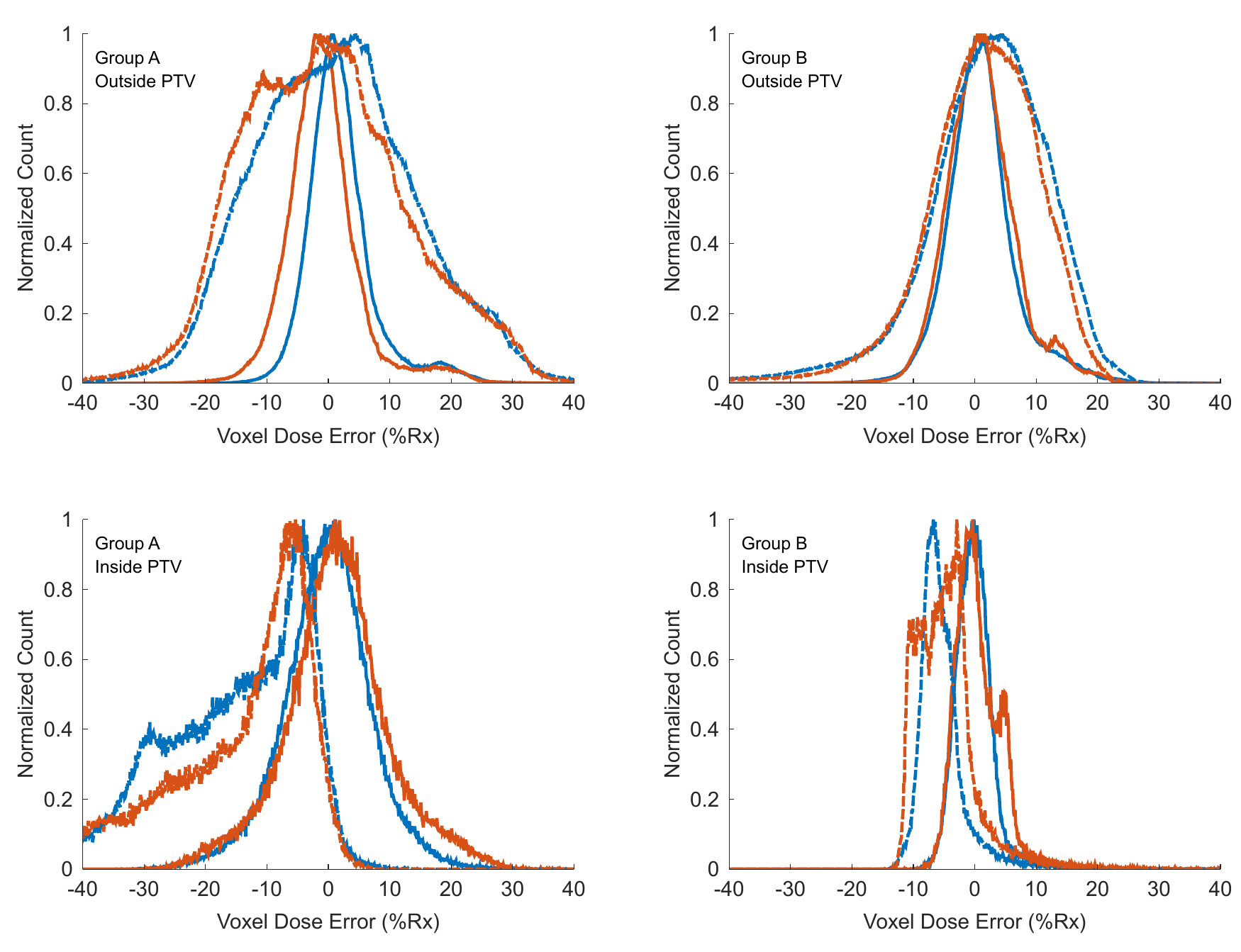}%
 \caption{Voxel dose errors as a percentage of the prescribed dose are shown for Groups A and B.  Top: errors from ANN-DMs are plotted alongside errors from simple linear regression models outside the PTV.  Bottom: errors from ANN-DMs are plotted alongside errors from the null model inside the PTV.}
 \end{figure*}

\subsection{Simple Dose Model Comparisons}
Voxel error distributions are plotted in Figure 6 for the final versions of each ANN-DM, the null model, and simple linear regression models.  For each, training and validation data are both shown.  Outside the PTV, dose distributions based on linear regression dose models demonstrated much broader error ranges than ANN-DM dose distributions.  Within the PTV, the null model was shown to underestimate the dose delivered, indicating that a noticeable region of the PTV receives more than the nominal prescribed dose in physician-approved clinical treatment plans.

\section{Discussion}

We have implemented a neural-network-based 3D dose-prediction algorithm, and have trained/validated it using 43 clinically approved pancreatic SBRT plans.  The algorithm is able to predict 3D dose distributions in the vicinity of pancreatic tumors with errors on the order of 5\%.  ANN-DMs were also shown to significantly outperform more simple models (i.e., null model and simple linear regression).  However, there remains substantial room for improvement of the models.  One advantage of neural network machine learning is that new information can be incorporated with relative ease by the additional of parameters to the model.  We note that, in Figure 3c and 3f, there is a substantial component of error related to angle, although the addition of a single angle (i.e., \enm{\theta}) parameter was insufficient to account for this difference.  One possibility is that the model may require additional information regarding the path of each radiation beamlet (e.g., whether there are overlying PTV or OAR voxels).

	Our results demonstrate the appropriateness of artificial neural networks for this purpose. Compared to more simple models (null models or linear regression with feature engineering), ANN-DM displayed greatly reduced error (see Figure 6).  The underlying structure of planned radiotherapy dose distributions is inherently multi-dimensional, and modeling these distributions requires a sophisticated approach.  Additionally, we found no benefit to more complex models (e.g. increasing the complexity of the neural network structure).  

As has been demonstrated, artificial neural network dose models can be used to reliably predict physician-approved dose distributions for patients receiving pancreatic SBRT.  Furthermore, the predictive accuracy of these dose models was significantly improved when dedicated models were developed for specific treatment styles.  As such, despite being implemented for the same treatment site using similar dose constraints, we have shown that neural networks are capable of adapting to different treatment protocols for the same site.  This specificity is important, because one motivating force behind knowledge-based planning is to somehow quantify and characterize the essential aspects of high-quality treatment plans.  Theoretically, one could obtain treatment plans from institutions boasting the best treatment outcomes, and train ANN-DMs to predict the optimal dose distributions for each new patient.  In addition to promoting high-quality radiation therapy, such dose models could be used to ensure adherence to treatment protocols in large-scale clinical trials.  Practically, an ANN-DM validation scheme could be implemented with relative ease, given that the model could be trained centrally by the trial coordinates and distributed to member institutions for local calculation and validation.

It should be noted that, although we have shown that ANN-DMs are capable of predicting desirable dose distributions, these models alone do not describe how to deliver these dose distributions.  Yet, because these models are based off of previously delivered treatment plans, they are unlikely to propose outrageously idealistic dose distributions.  ANN-DMs trained using real plans should produce deliverable or near-deliverable dose distributions.  Hence, one could envision a clinical workflow where ANN-DMs are used initially to predict a physician-approved dose distribution for each patient.  With this initial head start, dosimetrists can then focus their efforts on refining each plan.

In addition, neural networks are not infallible, and should not be used to autonomously guide treatment.  For example, neural networks can be susceptible to extrapolation errors whenever a new case is presented that resides outside the range of the training dataset (e.g., a PTV larger or smaller than was seen during training).  Thus, as optimal plans for outlier cases accumulate, they can be used to further train the model, thereby expanding the range of values the network is proficient in.  Another drawback of ANN-DM is that they do not directly calculate (and are not trained based on) DVH parameters, which are the most clinically relevant.  The ANN-DMs developed here were trained according to the dose errors seen for each voxel.  On the other hand, DVH errors are by their nature based off of cohorts of many voxels.  However, modifications to the current approach might improve DVH accuracy.  Theoretically, one could define new input parameters that might better represent the relationship between the PTV and OARs.  For example, inversely optimized arc-based treatment plans are unlikely to deliver dose through portions of the arc where the PTV and OARs are in line with one another.  Ergo, as determined by the geometries of the PTV, the OARs, and the treatment machine, one could delineate regions of patient anatomy that are less likely to be irradiated.  With these regions defined, a Boolean geometric parameter could be used to indicate whether or not a given voxel was in a region that was unlikely to be irradiated.

\section{Conclusions}

Artificial neural network dose models have been developed to reliably predict physician-approved dose distributions for pancreatic SBRT patients.  The influence of different neural network features, including network complexity and input parameter choices, have been investigated and networks have been optimized accordingly.  It is particularly noteworthy that significant predictive accuracy was gained by building separate ANN-DMs for separate treatment protocols.  As such, using consensus guidelines for high-quality treatment plans, neural network dose models could potentially be trained for the purposes of patient-specific plan quality validation.  The inclusion of such a validation step could help promote protocol compliance, be it for a single institution, or for a large-scale clinical trial.

\section{Acknowledgements}
This work was funded in part by the National Institutes of Health under award number K12CA086913, the University of Colorado Cancer Center/ACS IRG \#57-001-53 from the American Cancer Society, the Boettcher Foundation, and Varian Medical Systems.  These funding sources had no involvement in the study design; in the collection, analysis and interpretation of data; in the writing of the manuscript; or in the decision to submit the manuscript for publication.

\section*{References}
{\small 
[1] Vincent A, Herman J, Schulick R, Hruban RH, Goggins M. Pancreatic cancer. Lancet. 2011;378:607-620.

[2] Kamisawa T, Wood LD, Itoi T, Takaori K. Pancreatic cancer. Lancet. 2016;388:73-85.

[3] Ceha HM, van Tienhoven G, Gouma DJ, et al. Feasibility and efficacy of high dose conformal radiotherapy for patients with locally advanced pancreatic carcinoma. Cancer. 2000;89(11):2222-2229.

[4] Chang DT, Schellenberg D, Shen J, et al. Stereotactic radiotherapy for unresectable adenocarcinoma of the pancreas. Cancer. 2009;115(3):665-672.

[5] Koong AC, Le QT, Ho A, et al. Phase I study of stereotactic radiosurgery in patients with locally advanced pancreatic cancer. Int J Radiat Oncol Biol Phys. 2004;58(4):1017-1021.

[6] Koong AC, Christofferson E, Le QT, et al. Phase II study to assess the efficacy of conventionally fractionated radiotherapy followed by a stereotactic radiosurgery boost in patients with locally advanced pancreatic cancer. Int J Radiat Oncol Biol Phys. 2005;63(2):320-323.

[7] Mahadevan A, Miksad R, Goldstein M, et al. Induction gemcitabine and stereotactic body radiotherapy for locally advanced nonmetastatic pancreas cancer. Int J Radiat Oncol Biol Phys. 2011;81(4):e615-e622.

[8] Chuong MD, Springett GM, Freilich JM, et al. Stereotactic body radiation therapy for locally advanced and borderline resectable pancreatic cancer is effective and well tolerated. Int J Radiat Oncol Biol Phys. 2013;86(3):516-522.

[9] Timmerman R, Paulus R, Galvin J, et al. Stereotactic body radiation therapy for inoperable early stage lung cancer. JAMA. 2010;303(11):1070-1076.

[10] Stinauer MA, Kavanagh BD, Schefter TE, et al. Stereotactic body radiation therapy for melanoma and renal cell carcinoma: impact of single fraction equivalent dose on local control. Radiat Oncol. 2011;6(34):1-8.

[11] Park HJ, Griffin RJ, Hui S, Levitt SH, Song CW. Radiation-induced vascular damage in tumors: implications of vascular damage in ablative hypofractionated radiotherapy (SBRT and SRS). Radiat Res. 2012;177(3):311-327.

[12] Whipple AO, Parsons WB, Mullins CR. Treatment of carcinoma of the ampulla of vater. Ann Surg. 1935;102(4):763-779.

[13] Schnelldorfer T, Sarr MG. Alessandro Codivilla and the first pancreatoduodenectomy. Arch Surg. 2009;144(12):1179-1184.

[14] Are C, Dhir M, Ravipati L. History of pancreaticoduodenectomy: early misconceptions, initial milestones and the pioneers. HPB. 2011;13(6):377-384.

[15] Peters JH, Carey LC. Historical review of pancreaticoduodenectomy. Am J Surg. 1991;161:219-225.

[16] Lewis R, Drebin JA, Callery MP, et al. A contemporary analysis of survival for resected pancreatic ductal adenocarcinoma. HPB. 2013;15(1):49-60.

[17] Morris PJ, Nardi GL. Pancreaticoduodenal cancer: experience from 1951 to 1960 with a look ahead and behind. Arch Surg. 1966;92(6):834-837.

[18] Cameron JL, Riall TS, Coleman J, Belcher KA. One thousand consecutive pancreaticoduodenectomies. Ann Surg. 2006;244(1):10-15.

[19] Birkmeyer JD, Finlayson S, Tosteson A. Effect of hospital volume on in-hospital mortality with pancreaticoduodenectomy. Surgery. 1999;125(3):250-256.

[20] Abrams RA, Winter KA, Regine WF, et al. Failure to adhere to protocol specified radiation therapy guidelines was associated with decreased survival in RTOG 9704–a phase III trial of adjuvant chemotherapy and chemoradiotherapy for patients with resected adenocarcinoma of the pancreas. Int J Radiat Oncol Biol Phys. 2012;82(2):809-816.

[21] Nelms BE, Robinson G, Markham J, et al. Variation in external beam treatment plan quality: an inter-institutional study of planners and planning systems. Radiat Oncol. 2012;2(4):296-305.

[22] Amini A, Jones BL, Ghosh D, Schefter TS, Goodman KA. Impact of facility volume on outcomes in patients with squamous cell carcinoma of the anal canal: analysis of the National Cancer Data Base. Cancer. 2017;123(2):228-236.

[23] Shiraishi S, Moore KL. Knowledge-based prediction of three-dimensional dose distributions for external beam radiotherapy. Med Phys. 2016;43(1):378-387.

[24] Jabbour SK, Hashem SA, Bosch W, et al. Upper abdominal normal organ contouring guidelines and atlas: a Radiation Therapy Oncology Group consensus. Pract Radiat Oncol. 2014;4(2):82-89.
}

% Create the reference section using BibTeX:
%\bibliography{your-bib-file}

\end{document}